\documentclass[jkps,preprint,fleqn,showpacs,showkeys]{revtex4}
\usepackage[pdftex]{graphicx}
\usepackage{amssymb}
\usepackage{amsmath}
\usepackage{bm}
\usepackage{caption}
\begin{document}
\setcounter{page}{1}
\title[]{Beam Characterization at the KAERI UED Beamline}
\author{Sadiq \surname{Setiniyaz}}
\email{sadik82@gmail.com}
\thanks{Fax: +82-42-866-6150}
\author{Hyun Woo \surname{Kim}}
\author{In-Hyung \surname{Baek}}
\author{Jinhee \surname{Nam}}
\author{MoonSik \surname{Chae}}
\author{Byung-Heon \surname{Han}}
\author{Boris \surname{Gudkov}}
\author{Kyu Ha \surname{Jang}}
\author{Sunjeong \surname{Park}}
\author{Sergey \surname{Miginsky}}
\author{Nikolay \surname{Vinokurov}}
\author{Young Uk \surname{Jeong}}
\affiliation{Korea Atomic Energy Research Institute, Daejeon, 989-111}
\author{Sergey \surname{Miginsky}}
\author{Nikolay \surname{Vinokurov}}
\affiliation{Budker Institute of Nuclear Physics Siberian Branch of Russian Acadamy of Science, Novosibirsk, Russia}

\date[]{Received 2 November 2015}

\begin{abstract}
The UED (ultrafast electron diffraction) beamline of the KAERI's (the Korea Atomic Energy Research Institute's) WCI (World Class Institute) Center has been successfully commissioned. We have measured the beam emittance by using the quadrupole scan technique and the charge by using a novel measurement system we have developed.
In the quadrupole scan, a larger drift distance between the quadrupole and the screen is preferred because it gives a better thin-lens approximation. A high bunch-charge beam, however, will undergo emittance growth in the long drift caused by the space-charge force. 
We present a method that mitigates this growth by introducing a quadrupole scan with a short drift and without using the thin-lens approximation. 
The quadrupole in this method is treated as a thick lens, and the emittance is extracted by using the thick-lens equations.
Apart from being precise, our method can be readily applied without making any change to the beamline and has no need for a big drift space.
For charge measurement, we have developed a system consisting of an in-air Faraday cup (FC) and a preamplifier. 
Tests performed utilizing 3.3-MeV electrons show that the system was able to measure bunches with pulse durations of tens of femtoseconds at 10 fC sensitivity.

\end{abstract}



\maketitle

\section{INTRODUCTION}

\begin{figure}
	\includegraphics*[width=85mm]{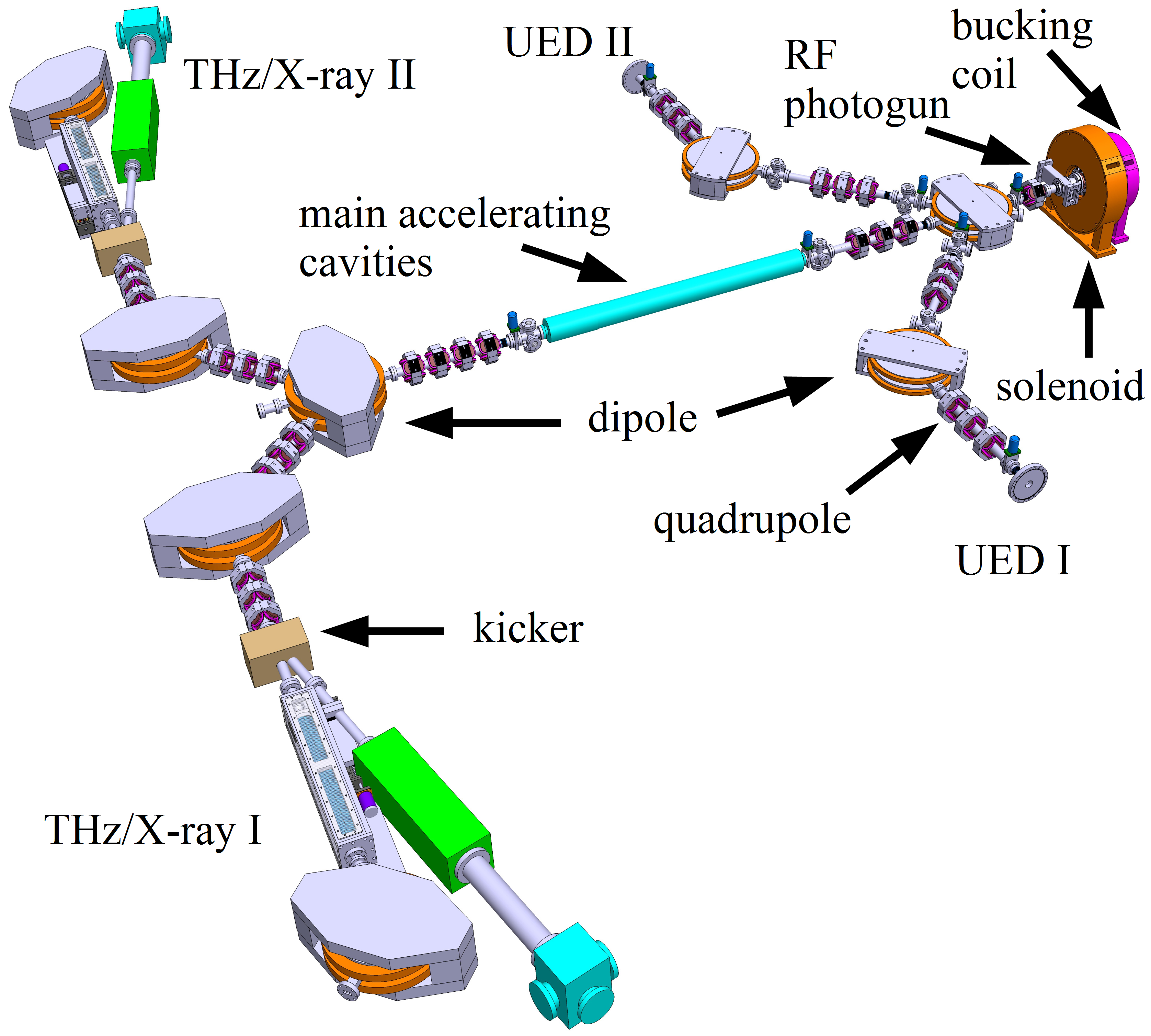}
	\caption{Layout of the electron beamline at KAERI's WCI center.}
	\label{wci-layout}
\end{figure}

The radio-frequency (RF) photogun of the World Class Institute (WCI) Center of the Korea Atomic Energy Research Institute (KAERI) is designed to generate sub-picosecond electron bunches with energies up to 3.3~MeV. The gun is a S-band co-axial RF photogun and has 1.5 cylindrically symmetric cells to remove multiple modes of the electric filed inside the cavity.
The electrical beam from the gun can be delivered to ultrafast electron diffraction (UED) experiments or can be further accelerated up to 20 - 30~MeV by using the main accelerating cavity for X-ray/THz pump and probe experiments, as shown in the Fig.~\ref{wci-layout}. The UED beamline supplies electron bunches with a 0.1-ps length, 1 to tens of pC charge, and a nominal energy of 3~MeV by utilizing an achromatic bend via velocity bunching~\cite{vinokurov}. We successfully commissioned this beamline and measured the electron beam parameters.

The transverse emittance and the Twiss parameters are important parameters of an accelerator to quantify the beam quality and match optics. The most common methods to measure the emittance are the quadrupole scan~\cite{emit-sadiq,Carlsten,Chiadroni,Spesyvtsev,CThomas}, the slit and collector~\cite{Steenbergen,Walter}, and the pepper-pot~\cite{Kubo} methods. The quadrupole scan is one of the simplest techniques to measure the emittance. In the quadrupole scan, the beam size is measured as a function of the quadrupole's magnetic field strength~\cite{emit-sadiq}. Imaging screens such as OTR (optical transition radiation), YAG (yttrium aluminum garnet), or phosphor screens are used along with a synchronized camera to observe the beam profile. Generally, the thin-lens approximation is applied and root mean square (rms) beam sizes obtained from the beam profile are used to extract the emittance and the Twiss parameters by fitting a parabolic function to the data. 

The thin-lens approximation is valid when $\sqrt{k_{1}}{L}<<1$, where $k_{1}$ is the quadrupole strength and $L$ is its effective length. The quadrupole here is viewed as a thin focusing/de-focusing lens. Therefore, $k_{1}$ is kept small while the drift distance between the quadrupole and the screen is set as large as possible (usually few meters). However, a high-bunch charge beam passing through a long drift will experience emittance growth due to the space charge force~\cite{Sawyer}. This growth, however, can be mitigated by shortening the drift length and, thus, extracting the emittance without using the thin-lens approximation. In our case, the drift length is shortened to 23~cm, i.e., one order of magnitude smaller than in the general case. In our method, the quadrupole is treated as a thick lens, and the emittance is extracted by using thick lens equations.

\section{Thick-Lens Quadrupole Scan}
In a quadrupole scan, a quadrupole magnet and a screen are used to measure the emittance and the Twiss parameters of the beam. The screen is separated from the quadrupole by a drift space. The transfer matrix of the scanning region $\mathbf{\mathbf{M}}$ is given by the matrix product of the transfer matrices of drift $\mathbf{\mathbf{S}}$ and quadrupole $\mathbf{\mathbf{Q}}$:
\begin{equation}
	\label{scan-Tmatrix}
	\mathbf{\mathbf{M}}=\mathbf{\mathbf{SQ}}=\Bigl(\begin{array}{cc}
		m_{11} &m_{12}\\
		m_{21} & m_{22}
	\end{array}\Bigr).
\end{equation}
\noindent Here, $\mathbf{\mathbf{S}}$ and $\mathbf{\mathbf{Q}}$ are defined as
\begin{equation}
\label{drift-Tmatrix}
\mathbf{\mathbf{S}}\equiv\Bigl(\begin{array}{cc}
1 & l\\
0 & 1
\end{array}\Bigr),
\end{equation}
\begin{equation}
	\label{quadrupole-Tmatrix}
	\mathrm{\mathbf{Q}}\equiv\Bigl(\begin{array}{cc}
		cos\sqrt{k_{1}}L & \frac{1}{\sqrt{k_{1}}}sin\sqrt{k_{1}}L\\
		-\sqrt{k_{1}}sin\sqrt{k_{1}}L & cos\sqrt{k_{1}}L
	\end{array}\Bigr),
\end{equation}

\noindent where $l$ is the drift length. The beam matrix at the screen $ \mathbf{\sigma_{s}}$ is related to the beam matrix at the quadrupole $\mathbf{\sigma_{q}}$ by using the similarity transformation as~\cite{SYLee}
\begin{equation}
	\mathbf{\mathbf{\sigma_{s}=M\mathrm{\mathbf{\mathbf{\sigma_{q}}}}}M}^{\mathrm{T}},
\end{equation}
with $\mathbf{\sigma_{s}}$ and $\mathbf{\sigma_{q}}$ being defined as
\begin{equation}
	\mathbf{\mathbf{\sigma_{s,\mathnormal{x}}}}\equiv\Bigl(\begin{array}{cc}
		\sigma_{\textnormal{s},x}^{2} & \sigma_{\textnormal{s},xx'} \\
		\sigma_{\textnormal{s},xx'} & \sigma_{\textnormal{s},x'}^{2}
	\end{array}\Bigr)
	,\; 
	\mathbf{\mathbf{\sigma_{q,\mathnormal{x}}}}\equiv\Bigl(\begin{array}{cc}
		\sigma_{\textnormal{q},x}^{2} & \sigma_{\textnormal{q},xx'}\\
		\sigma_{\textnormal{q},xx'}   & \sigma_{\textnormal{q},x'}^{2}
	\end{array}\Bigr).
\end{equation}
\noindent The matrix element $\sigma_{\textnormal{s},x}$/$\sigma_{\textnormal{q},x}$ is the horizontal rms beam size at the screen/quadrupole. $\sigma_{\textnormal{s},x}$ can be expressed as a function of the transfer matrix elements $m_{11}$ and $m_{12}$ as
\begin{equation}
	\sigma_{\textnormal{s},x}^{2}=\sigma_{\textnormal{q},x}\beta_{\textnormal{q},x} \Bigl( m_{11}+m_{12}\frac{-\alpha_{\textnormal{q},x}}{\beta_{\textnormal{q},x}} \Bigr)^2 + m_{12}^2 \frac{\sigma_{q,x}}{\beta_{q,x}},
	\label{fit}
\end{equation}
\noindent where  $m_{11}$ and $m_{12}$ are given by
\begin{equation}
	\label{m11-m12}
	\begin{array}{c}
		m_{11}=cos\sqrt{k_{1}}L - l\sqrt{k_{1}}sin\sqrt{k_{1}}L,\\ \\
		m_{12}=\frac{1}{\sqrt{k_{1}}}sin\sqrt{k_{1}}L + lcos\sqrt{k_{1}}L.
	\end{array} 
\end{equation}

When the thin-lens approximation is valid, Eq.~(\ref{fit}) becomes a parabolic function. The emittance and the Twiss parameters are extracted by measuring $\sigma_{\textnormal{s},x}$ and fitting a parabolic function to the data. Because we used a short drift distance to mitigate emittance growth, the thin-lens condition $k_{1}L<<1$ is no longer valid. Thus, we shall obtain the emittance and the Twiss parameters by directly fitting Eq.~(\ref{fit}) and treating the quadrupole as a thick lens. 

\section{Emittance Measurement}
\subsection{Experiment Setup}

\begin{figure}
	\includegraphics*[width=75mm]{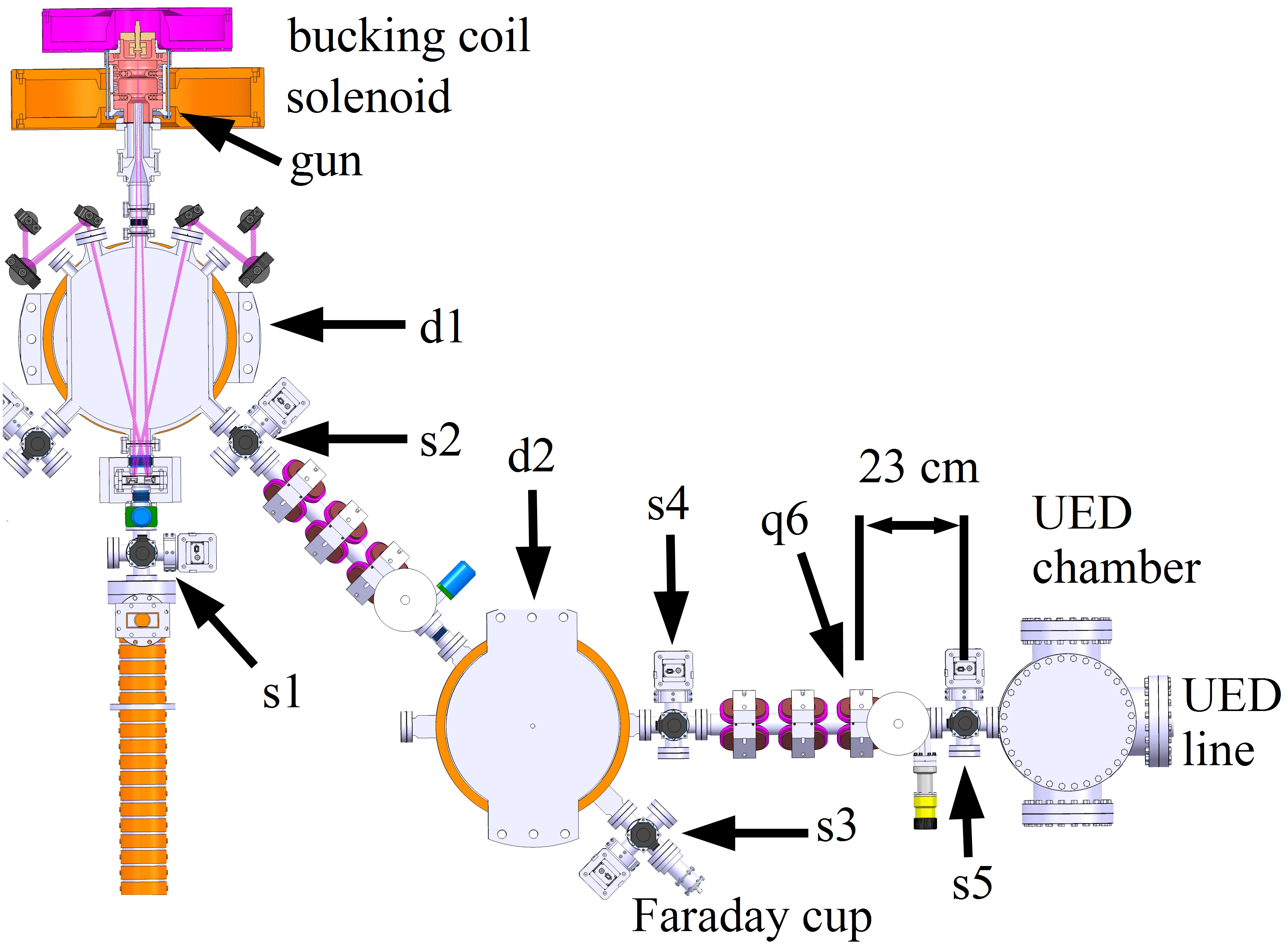}
	\caption{Experiment setup for the quadruple scan.}
	\label{emit_setup}
\end{figure}

The experimental setup for the emittance measurement with the quadrupole scan is shown in Fig.~\ref{emit_setup}. The electron beam from the RF photogun is delivered to the UED chamber by using two 45$^{\textnormal{o}}$ dipole magnets and 6 quadrupoles. Five retractable imaging screens, s1~-~s5, are used to observe the beam. The coil current of quadrupole q6 is varied to perform scanning. A P-22 type phosphor screen, s5, with a 12.7-mm diameter is located 23.0~cm downstream of the q6. A synchronized camera (Basler scout scA 600-28fm) placed under the screen is used to observe the beam profile. 

The beam energy and its spread are measured by using the dipole d1 and the screen s2. The magnetic field of the d1 is mapped to determine the energy of the beam when it is bent by 45$^{\textnormal{o}}$. A Faraday cup (FC) is placed at the end of the 45$^{\textnormal{o}}$ line to measure the electron-bunch charge. The effective length of the quadrupole was measured to be 8.335~cm by mapping its magnetic field.

\subsection{Experiment Procedure}
As shown in the Fig.~\ref{emit_setup}, the electrons in the gun are first focused by the solenoids. The beam out of the gun is observed at s1 screen and to tune a round shape. Then, it is bent 45$^{\textnormal{o}}$ by d1 and observed on s2. By observing the beam position on s2, we can maximize the beam energy by adjusting the laser delay. Then, d1 is turned off, and the beam is observed on s1 and tuned to a round shape again. Then, the beam is bent again by d1 and observed on s2. The coil current of d1 is recorded to estimate the beam energy and its spread. After this, the beam is delivered to s3 by using d1 and q1-q3, where it is centered and tuned to a round shape. The screen s3 is then retracted from the beamline, and the beam is dumped to the FC where it creates an electrical signal. This signal is too weak to be directly measured and thus is amplified by using a preamplifier, after which it can be observed using an oscilloscope. The upstream magnets are tuned to maximize the peak voltage of the signal, which is linearly related to the electron-bunch charge. 

After the charge measurement is finished, d2 is turned on, and the beam is observed on s4 at first and on s5 later. The horizontal beam size at s5 usually is larger than the vertical one because of beam dispersion. The dispersion is suppressed by adjusting q1 - q3 while q4 - q5 are turned off. When dispersion is suppressed, q6 should be able to focus the beam into a narrow horizontal shape on s5. After dispersion is suppressed, q4 and q5 are tuned to form a round beam on s5 while q6 is off. The scan is performed by incrementally changing the coil current of q6 and recording the corresponding beam image by using a camera.

\section{Charge Measurement System}
Faraday cups have long been used to measure the beam charge in particle accelerators~\cite{Dadourian, Allen, Trump, Stier, Allison, Williams, Welsh}. The beam stopper of the FC is made of conductive materials such as carbon or metals to collect the beam charge and transmit electrical signals. It must be designed to stop the beam, suppress secondary electron emission and recapture back scattered electrons. The shape and the material of the FC may differ depending on the beam conditions and applications~\cite{Cambriay, Seamans, Morain, Elmer} and can be optimized by simulating the beam interaction with the stopper~\cite{Thomas, Kashefian}. 

Matching the impedance between the FC, preamplifier and test network is another important design consideration~\cite{Hu}. The rise time and the amplitude of the electrical signal from the FC depends on the incident-particle beam current. The KAERI UED beamline is designed to generate electron bunches with a 10- to 100-A peak current~\cite{vinokurov}, which correspond to maximum signal frequencies up to tens of THz. Measuring this charge with high sensitivity is challenging because of the ultrashort bunch length, high frequency, low charge, and noises from RF power source. The idea we are proposing to solve this problem is to use a charge-sensitive preamplifier near the FC to integrate, amplify, and convert this current signal to a voltage signal that can be measured via an oscilloscope or ADC at remote locations. We have fabricated, calibrated, and tested this preamplifier by using an ultrashort electron beam and were able to achieve a 10-fC precision. 

In our setup, electron bunches passing through a 0.5-mm-thick aluminum vacuum window are deposited into an in-air FC 5~mm away. The FC is shielded with stainless steel from external noise and can be attached to/removed from a 2.75-inch flange by using tapped holes without breaking the vacuum. To optimize the FC geometry and increase the detection efficiency, we simulated the interactions of electrons with the vacuum window and the aluminum FC stopper by using the G4beamline~\cite{Muons}. 

\begin{figure}
	\includegraphics*[width=75mm]{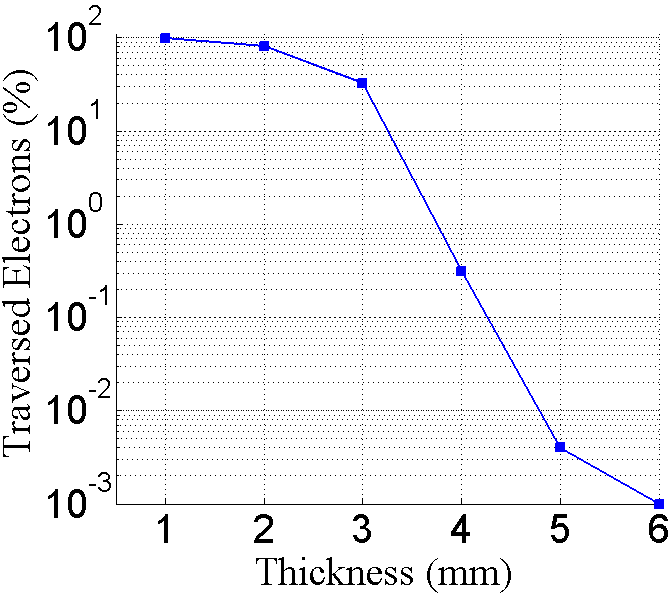}
	\caption{Percentage of electrons traversing different thicknesses of the aluminum base (without a side wall).}
	\label{FCBL_vs_Eff}
\end{figure}

The first step in the optimization is to find the proper thickness of the FC stopper base (without side wall) that would stop most of the 3-MeV electrons. The simulations were performed for aluminum with a 25-mm diameter while the thickness was varied from 1 to 6 mm in 1-mm increments, and the results are in Fig.~\ref{FCBL_vs_Eff}. The FC base and the vacuum window are placed 5~mm apart. When the aluminum base is 5-mm thick, less than 0.01$\%$ of the electrons were able to pass through. Therefore, the minimum thickness of the base and side wall is decided to be 5~mm. Because the stopper is mounted to the main body of the FC via a M6 bolt (stainless steel), an additional 5~mm was added to the base for mechanical attachment, which makes the base 10~mm thick. 

\begin{figure}
	\includegraphics*[width=75mm]{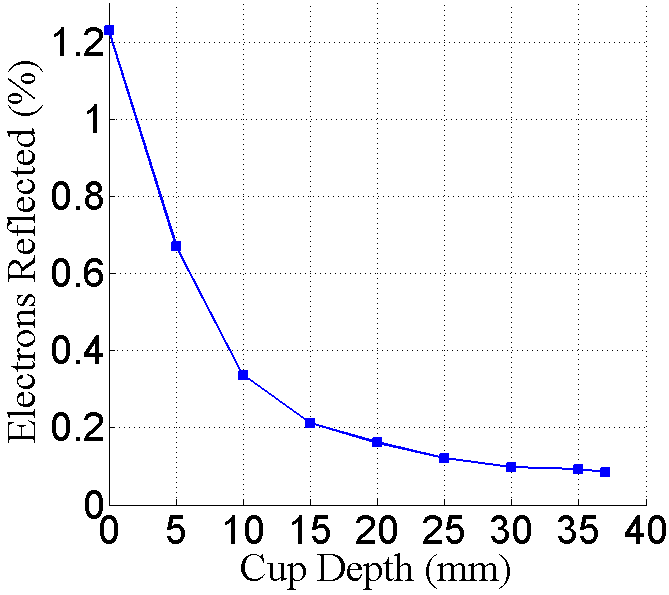}
	\caption{Percentage of electrons reflected by different cup depth. The base and side wall thicknesses are 10~mm and 5~mm respectively.}
	\label{FCSL_vs_Eff}
\end{figure}

\begin{figure}
	\includegraphics*[width=90mm]{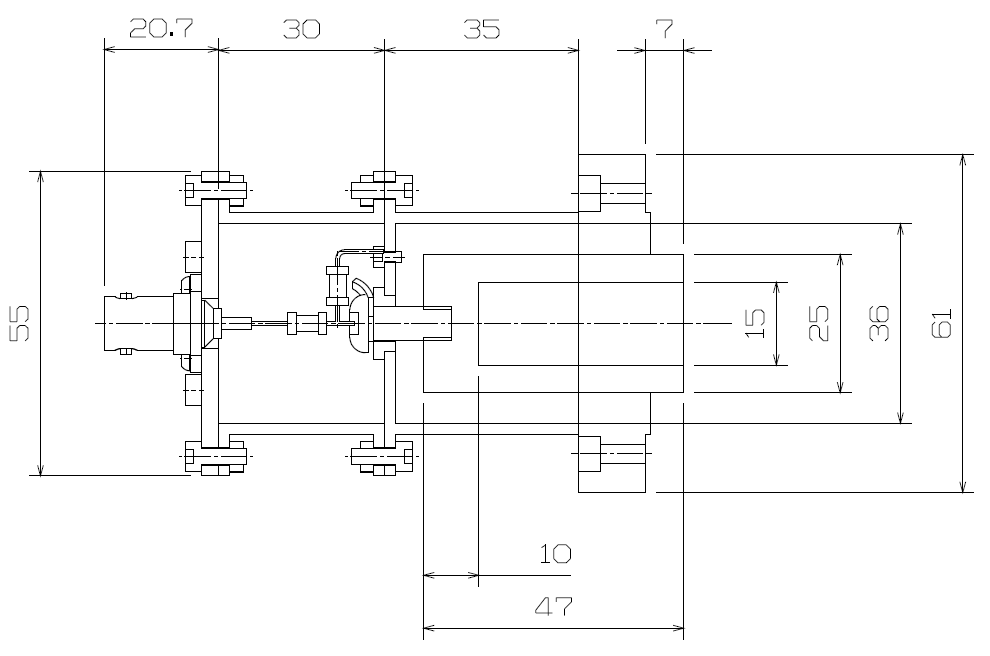}
	\caption{Optimized FC dimension (in mm) for the 3-MeV electron beam.}
	\label{FC}
\end{figure}

Electrons are also lost due to reflection by the metal surface. The next step in the simulation, therefore, is to determine the stopper's side wall length (i.e., cup depth) and thickness. A 5-mm wall thickness is sufficient, as mentioned above. The simulations were carried out for 0- to 37-mm cup depths, and the results are shown in the Fig.~\ref{FCSL_vs_Eff}. The inner diameter of the cup is 15~mm. When the cup depth is 37~mm, less than 0.1$\%$ of the electrons were reflected back. With the optimized dimensions of the FC, as shown in Fig.~\ref{FC}, over 99.8\% of the 3-MeV electrons were captured.

\begin{figure}
	\includegraphics*[width=150mm]{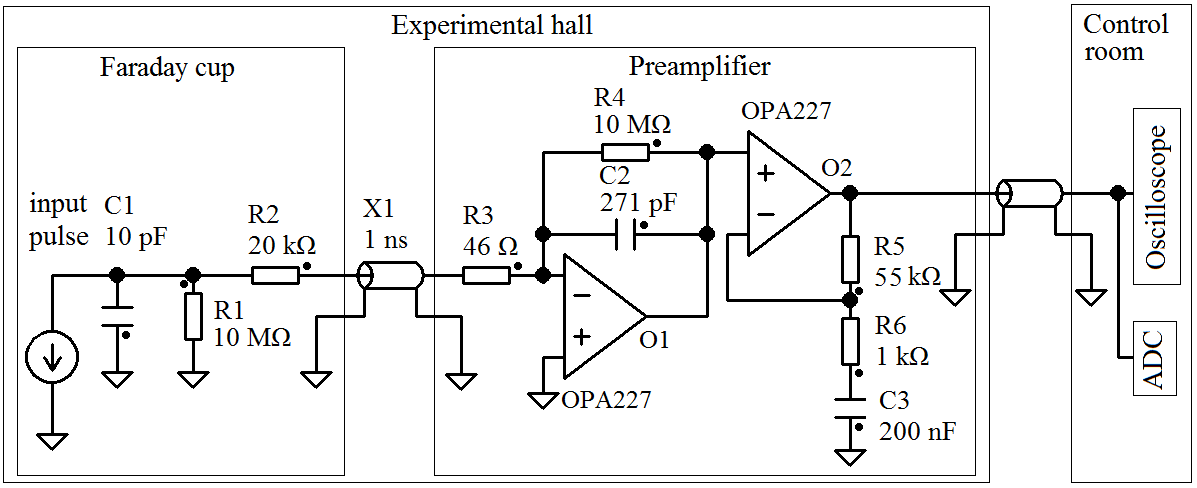}
	\caption{The FC, preamplifier, and measurement network circuits.}
	\label{FC2}
\end{figure}

The M6 bolt is electrically isolated from the external ground shield and is connected to the 10-M$\Omega$ and 20-k$\Omega$ resistors. The 20-k$\Omega$ resistor is connected to a BNC connector which connects the FC to the preamplifier. The deposited charge creates a voltage signal across the FC stopper and external ground shield, which is transmitted to the preamplifier. The transmission and the amplification of the signal in the FC and preamplifier were simulated by using NL5~\cite{NL5}. The electrical circuit diagram for the FC and preamplifier is shown in Fig.~\ref{FC2}. The preamplifier consists of two op-amp instrumentation amplifiers. The capacitor C1 represents the capacitance between the FC stopper and the external grounded shield. C1 is estimated to be around 10~pF. A-0.2-m-long BNC cable X1 with a 20-pF capacitance connects the FC to the preamplifier. 
The preamplifier is powered by using four 8-volt batteries, has a linear response up to 16~V, and can measure charge up to 80~pC. One can also change the calibration factor (i.e., voltage-to-charge ratio) and increase/decrease the measurable charge limit by changing the feedback capacitor C2.

\begin{figure}
	\includegraphics*[width=75mm]{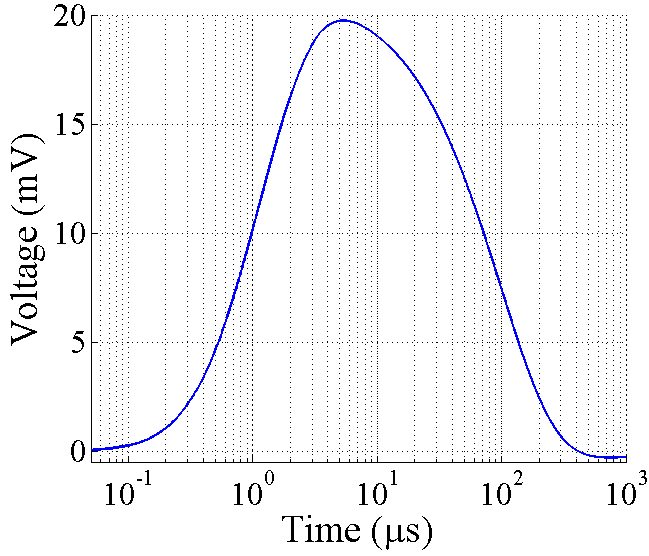}
	\caption{FC and preamplifier NL5 circuit simulation result. The  curve is the output voltage of the preamplifier (O2 in Fig.~\ref{FC2}) when a 0.1-pC charge is deposited to capacitor C1.}
	\label{PreamVvsT}
\end{figure}

In the simulation, a 0.1-pC charge is created by the input pulse and deposited to capacitor C1. This charge creates a voltage at C1, which is then amplified by using the preamplifier. The output pulse from the preamplifier is shown in Fig.~\ref{PreamVvsT}. The peak voltage, 19.8~mV, occurs about 5~$\mu$s after the input pulse; thus, the voltage-to-charge ratio of the preamplifier is 198~mV/pC. The output signals returns to zero within 1~ms, and this allows the preamplifier to operate at a 1-kHz repetition rate.

Experimentally, the preamplifier is calibrated by using a power supply to deposit charge to the FC stopper, and the result is
 \begin{equation}
 	\textnormal{Q(pC)} = (-0.005 \pm 0.0002)+ (0.1923 \pm 0.0003) \textnormal{U(V)},
 \end{equation}
where Q is the bunch charge deposited in the unit of pC, and U is the peak voltage from preamplifier, in volts, measured by using an oscilloscope. The experimentally-measured voltage-to-charge ratio of 192~mV/pC is similar to the simulation prediction of 198~mV/pC (off by 3$\%$).
\section{Data Analysis and Results}
\begin{figure}
	\includegraphics*[width=75mm]{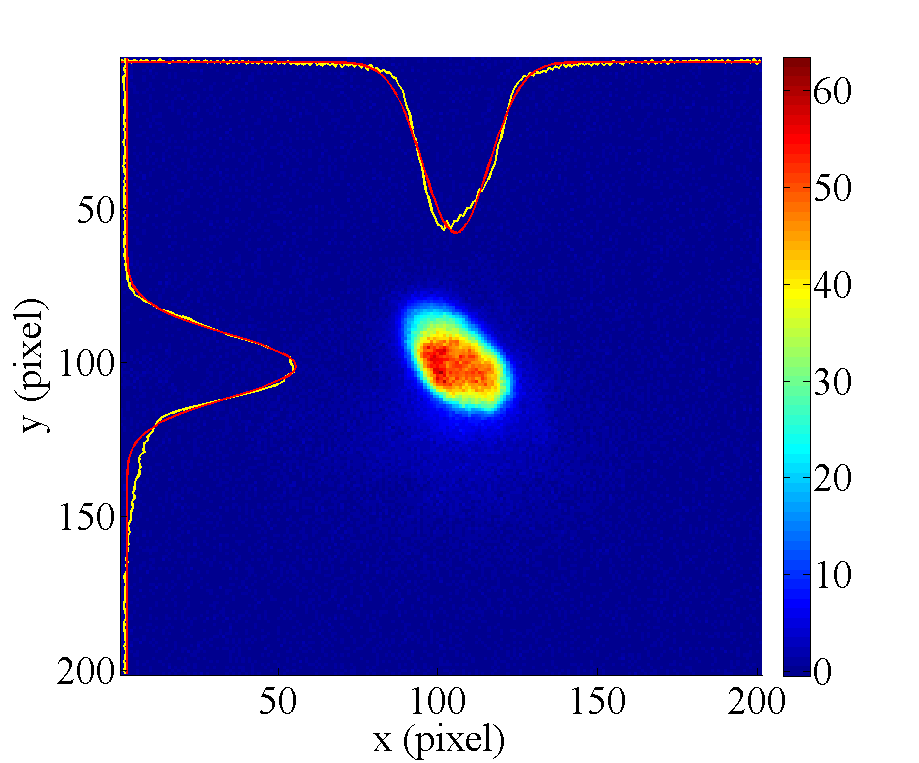}
	\caption{Electron beam profile observed on a phosphor screen when q6 is off. The yellow/red line is a projection/Gaussian fit.}
	\label{beam}
\end{figure}

Beam images from the camera were calibrated by using the screen target's frame which has a 12.7-mm diameter. A scaling factor of 0.0158~$\pm$~0.0006~mm/pixel is obtained by dividing the diameter of the frame by the number of pixels in the diameter. A MATLAB-based script was used to process the beam images and to calculate the emittance. The beam profile observed on the phosphor screen is shown in Fig.~\ref{beam}. The yellow lines are beam projections and are fitted with a Gaussian distribution as shown by the red curve. 

\begin{figure}
	\begin{tabular}{cc}
		{\scalebox{0.4} [0.4]{\includegraphics{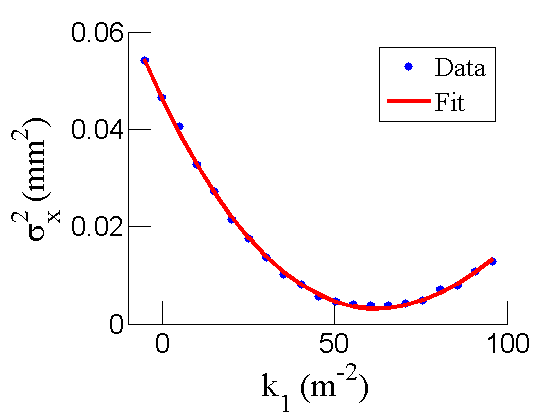}}} & {\scalebox{0.4} [0.4]{\includegraphics{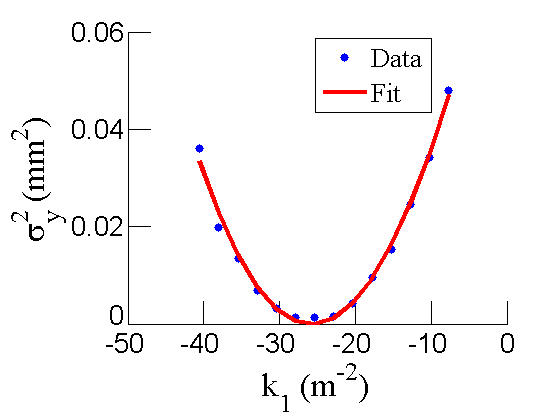}}}\\
		(a) & (b)\\	
	\end{tabular}
	\caption{Squares of rms beam size $vs$ $k_{1}$ and fit using the thick-lens equation: (a) horizontal beam projection, (b) vertical beam projection.}
	\label{fits}
\end{figure}

The emittance measurement was performed by changing the quadrupole current incrementally, which changes the quadrupole strength $k_{1}$, and the measuring the corresponding beam image on the viewing screen. The measured two-dimensional beam image was projected along the image's abscissa and ordinate axes. The rms value of the beam is extracted by fitting a Gaussian distribution to the beam projection. 
Measurements of $\sigma_\textnormal{s}$ for several quadrupole strengths ($k_{1}$) is then fit by using the function in Eq.~(\ref{fit}) to determine the emittance and the Twiss parameters.
Figure~\ref{fits} shows the square of the rms beam size ($\sigma^2_\textnormal{s}$) $vs$ $k_{1}$ for the $x$ (horizontal) and the $y$ (vertical) beam projections along with the fits obtained by using Eq.~(\ref{fit}). The emittances and the Twiss parameters from these fits are summarized in Table.~\ref{results} along with the results for the charge and the energy measurements.

\begin{table}
	\caption{Emittance measurement results.}
	\begin{ruledtabular}
		\begin{tabular}{ccc}
	{Parameter} & {Unit}  & {Value}    \\		
	normalized emittance $\epsilon_\textnormal{n,x}$ 	& mm$\cdot$mrad &   $0.23\pm0.03$ \\
	normalized emittance $\epsilon_\textnormal{n,y}$ 	& mm$\cdot$mrad &   $0.42\pm0.07$ \\
	$\beta_\textnormal{x}$-function                     &   m         &   $0.63\pm0.10$ \\
	$\beta_\textnormal{y}$-function                     &   m         &   $5.5 \pm1.6$  \\
	$\alpha_\textnormal{x}$-function                    &  rad        &   $0.9 \pm 1.1$ \\
	$\alpha_\textnormal{y}$-function                    &  rad        &   $0.23\pm0.03$ \\
	bunch charge                   		       		 	&  pC         &   $1.12  \pm  0.03$\\
	total energy $E$                 					&  MeV        &   3.26 \\
	relative energy spread $\Delta E/E$       			&  \%         &   0.96  \\		
		\end{tabular}
	\end{ruledtabular}
	\label{results}
\end{table}

\section{Summary and Discussion}
We have used an improved quadrupole scan method that utilizes a short drift distance and the thick-lens equation to measure the beam emittance of the KAERI's WCI center's UED linac. The horizontal/vertical emittances were measured to be $0.23 \pm 0.03$/$0.42 \pm 0.07$~mm$\cdot$mrad for a bunch charge of $1.12\pm0.03$~pC. The short drift distance not only reduces the growth of the emittance due to the space charge but also simplifies the experimental setup.

The beam energy and the relative energy spread were measured to be 3.26~MeV and 0.96\% respectively. The diffraction pattern is blurred when energy spread is present and according to the results given in Ref. 28, for a beam with a 0.96\% energy spread and a kinetic energy of 2.5 - 5.4~MeV, the blurring of the diffraction pattern is approximately 0.16 - 0.19. Here, blurring is obtained by normalizing the full width at half maximum (FWHM) of the 2$^{nd}$ ring to the gap between the 1$^{st}$ ring and the 2$^{nd}$ ring. With the current beam condition, we expect to observe  distinguishable diffraction patterns. Nevertheless, the energy spread can be further reduced by tuning the laser and/or using an energy slit.

We have developed a novel charge measurement system optimized for a 3-MeV beam with a pulse duration of tens of femtoseconds. The system demonstrated 10~fC sensitivity and can be further improved by reducing background noises and by replacing the smaller feedback capacitor and other electrical components of the preamplifier. Also, the measurable charge range can be increased by increasing the feedback capacitor's capacitance or some other methods. We plan to perform emittance $vs.$ charge measurements in the near future to understand the performance of our RF photogun.

\begin{acknowledgments}
This work was supported by the World Class Institute (WCI) Program of the National Research Foundation of Korea (NRF) funded by the Ministry of Science, ICT and Future Planning (NRF Grant No. WCI 2011-001).
\end{acknowledgments}

\end{document}